\documentclass[superscriptaddress,twocolumn,showpacs]{revtex4}

\usepackage{graphicx}
\usepackage{color}
\usepackage{dcolumn}
\usepackage{amsmath}
\usepackage{dsfont}

\DeclareMathOperator{\sech}{sech}

\begin{document}

\title{Spatial storage of discrete dark solitons}

\author{Alejadro J. Mart\'inez}
\affiliation{Oxford Centre for Industrial and Applied Mathematics,
Mathematical Institute, University of Oxford,\\Oxford OX2 6GG, United
Kingdom}

\author{Yair Z\'arate}
\affiliation{Nonlinear Physics Center, Research School of Physics and
Engineering, Australian National University, \\Canberra, Australian
Capital Territory 0200, Australia}


\begin{abstract}
The interaction between a mobile discrete dark soliton (DDS) and impurities in one-dimensional nonlinear (Kerr) photonic lattices is studied. We found that
the scattering is an inelastic process where the DDS can be reflected or transmitted depending on its transversal speed and the strength of the impurities.
In particular, in the reflection regime, the DDS increases its transversal speed after each scattering. A method for spatial storage of DDS solutions
using two impurities is discussed, where the soliton can be trapped within a storage region until it reaches the critical speed needed to be transmitted. We show numerically, that this method allows the storage of multiple DDSs simultaneously.
\end{abstract}
\maketitle

\section{Introduction}

Scattering of traveling waves by local impurities has
been a subject of extensive research, both theoretically and
experimentally, in a broad spectrum of physics~\cite{scattering,fano}.
In the framework of optics, this interaction provides the ability to control
localized traveling pulses by means of switching mechanisms. These can depend on, for example, the power of the pulse~\cite{power}, the momentum~\cite{k},
or even the
particular structure of the impurities~\cite{structural,trap}. Most of the research in this field has focused on the scattering of bright waves. Especially those
concerning discrete optical systems, where there is a high flexibility to construct impurities by manipulating the relative refraction index between the impurities
and the rest of the system~\cite{neg}. In particular, it is possible to generate either positive
or negative impurities with different resonance response~\cite{negdef}.
Different impurities permit to control transmission, reflection or even the localization degree of energy~\cite{trap2}.
It is interesting to note that these phenomena are analogous to the scattering of matter-waves by local
impurities studied in Bose-Einstein condensates~\cite{bec1,bec2,bec3}.
On the other hand, less attention has been given to the scattering of dark waves,
namely discrete dark solitons (DDS), with local impurities. Even though there is a considerable
amount of theoretical research on the subject (see e.g.~\cite{dds,dark1,dark2})
and DDS waves have been experimentally
observed for more than a decade~\cite{dexp1,dexp2}.

It is the aim of this article to study the scattering process of
a mobile DDS with local linear impurities. First we show that the interaction is an inelastic
process, where the DDS increases its transversal speed after the scattering. Then, following a continuous
approach based on adiabatic perturbation theory, we develop an
inelastic analysis for the position of the core of the DDS as function
of the propagation coordinate. This analysis allows us to obtain the critical (minimal) initial speed that a DDS requires
in order to be transmitted by a delta impurity. Applying our results to a system with two impurities, we show that the DDS can be stored between the
impurities for a certain propagation distance depending on the
initial speed and the strength of the impurities. Along the storage,
the DDS is successively reflected by the impurities until it reaches or
exceeds the critical speed to escape the confinement. We verify numerically that the proposed storage mechanism applies for either self-focusing or
self-defocusing systems. Where the resulting propagative nonlinear waves, except for the phase, exhibit the same behavior.
Finally, the storage for multiple DDS is reviewed. For symmetrical initial conditions, relative to the position of the impurities, the DDSs always escape in pairs.

\section{Model}
In the coupled modes framework, the propagation of light
along a nonlinear single-mode waveguide array can be described
by $\Psi(x,\xi)=\sum_n A_n(\xi)\Phi_n(x-na)$, where $\xi$ is the propagation
direction, $x$ is the spatial transversal coordinate and $a$ corresponds to the
separation between guides. The field $\Phi_n(x)$ is the mode of the $n$-th waveguide and the complex amplitude $A_n$ is described by the discrete nonlinear 
Schr\"odinger (DNLS) equation~\cite{GKBook},
\begin{eqnarray}
\label{DNLSwdim}
-i\frac{dA_n}{d\xi}&=&V(A_{n+1}+A_{n-1})+\left(\beta+\sum_{l=1}^M\delta_{n,n_l}\bar{\epsilon}\right)A_n\nonumber\\&& +\bar{\gamma}|A_n|^2A_n\,,
\end{eqnarray}
where $\beta$ is the linear propagation constant and $\delta_{n,n_l}\bar{\epsilon}$ represents the local perturbation
to the refractive index, i.e.~an impurity, located in the $n_l$-th waveguide. The parameter $\bar{\epsilon}$ accounts for the strength of the
impurity and it can take positive or negative values. $V$ is the coupling coefficient and $\bar{\gamma}=(\omega_0\bar{n}_2)/(cA_{eff})$ is the nonlinear parameter; 
with $\bar{\gamma}>0$ for the self-focusing case and $\bar{\gamma}<0$ for the 
self-defocusing case. Here, $\omega_0$ is the optical frequency associated with the modes, $\bar{n}_2$ is the Kerr coefficient, $c$ is the speed of light and $A_{eff}$ is the effective area 
of the waveguide mode \cite{GKBook}. Equation~\eqref{DNLSwdim} can be rescaled in order to reduce the number of parameters by considering $A_n(\xi)=\sqrt{V/\bar{\gamma}}\,u_n(z)e^{i\beta z/V}$, 
with $\xi=Vz$. Thus, Eq.~\eqref{DNLSwdim} now reads
\begin{equation}
\label{eq-DNLS}
-i\frac{du_n}{dz}=(u_{n+1}+u_{n-1})+\sum_{l=1}^M\delta_{n,n_l}\epsilon
u_n +\gamma|u_n|^2u_n\,,
\end{equation}
with $\epsilon \equiv \bar{\epsilon}/V$ and $\gamma \equiv \text{sgn}(\bar{\gamma})$, where $\text{sgn}$ is the signum function.
An important feature of Eq.~\eqref{eq-DNLS} is that it remains invariant under {\it staggered-unstaggered}
transformation: $\{z,\epsilon,\gamma,u_n\}\rightarrow\{-z,-\epsilon,-\gamma,(-1)^nu_n\}$. Hence, by engineering the appropriate phase, both nonlinear
regimes are equivalent. In fact, the balance between nonlinearity, dispersion and discreteness required by the existence of DDS can be achieved
in both cases~\cite{dds}. This leads to an unstaggered (staggered) wave for the self-defocussing (self-focusing) regime. Thus, without loss of generality,
we will focus our study on the self-defocusing case, with $\gamma =-1$. It is worth mentioning that a mobile DDS propagating in a self-defocusing
system without impurities can be described by $u_n=A\tanh(A(n-n_0))+iB$, where $\dot{n}_0\equiv\frac{dn_0}{dz}\approx B$ is the transversal speed of
the DDS \cite{dds}. For the sake of simplicity, we fix the background field as $A^2+B^2=1$, reducing the parameter space to $\{B,\epsilon\}$.

\begin{figure}[t!]
\centering
\includegraphics[width=8.5cm]{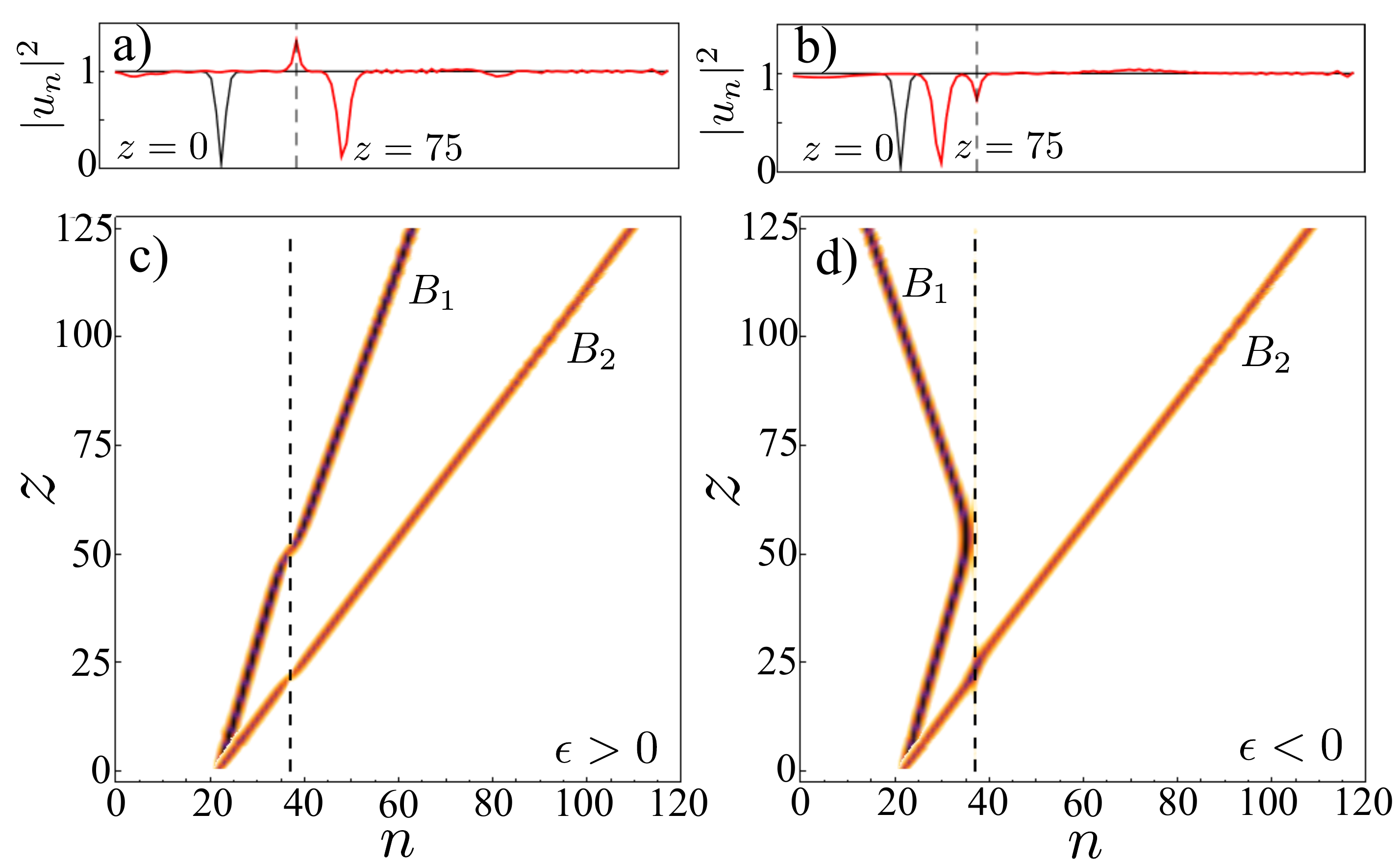}
\caption{Examples of scattering of DDSs with an impurity in a self-defocusing nonlinear media. a) and b) show the intensity distributions as a function of $n$ at $z=0$
(black) and $z=75$ (red), for a DDS with initial speed $B_1 = 0.2$. c) and d) show the dynamic associated with two independent initial
conditions in each case, with $B_2 = 0.5$. The vertical dashed lines denote the position of the impurity. The values of $\epsilon$ are $0.5$ for a) and c), and $-0.5$ for b) and d) panels.}
\label{fig-diffdyn}
\end{figure}
We conducted numerical simulations of a DDS in the nonlinear waveguide array in order to characterize the scattering process due to the
presence of
impurities. Equation ~\eqref{eq-DNLS} was integrated with a fourth-order Runge-Kutta scheme with a fixed-step size of $\delta z=10^{-3}$, this is in order to conserve the total power and 
the energy with a relative error lower than $10^{-6}$. 
To consider disturbances that emerge naturally in an experiment, we set as a initial condition, the
DDS profile for the homogeneous system (without impurities) as described above. Thus, once the DDS
begins to propagate, the energy is split among the DDS itself, fluctuations in the
background field and localized impurity modes [see
Figs.~\ref{fig-diffdyn}a and \ref{fig-diffdyn}b]. The magnitude of the background fluctuations is typically lower than 10\% of the amplitude of the background field, and in most of the cases
it is smaller than the amplitude of the localized impurity mode. 

For a single
positive impurity, i.e. $\epsilon>0$ and $M=1$ in Eq.~\eqref{eq-DNLS}, the DDS is always transmitted independently of the relation between the parameters
[see Fig.~\ref{fig-diffdyn}c]. Whereas, for a negative impurity, the DDS can be either reflected or transmitted [see Fig.~\ref{fig-diffdyn}d].
The transition depends on
the relation between the transversal speed of the DDS and the strength of the impurity. The remainder of this article will focus on the negative
impurities case ($\epsilon<0$), unless otherwise stated.
\begin{figure}[t]
\centering
\includegraphics[width=7cm]{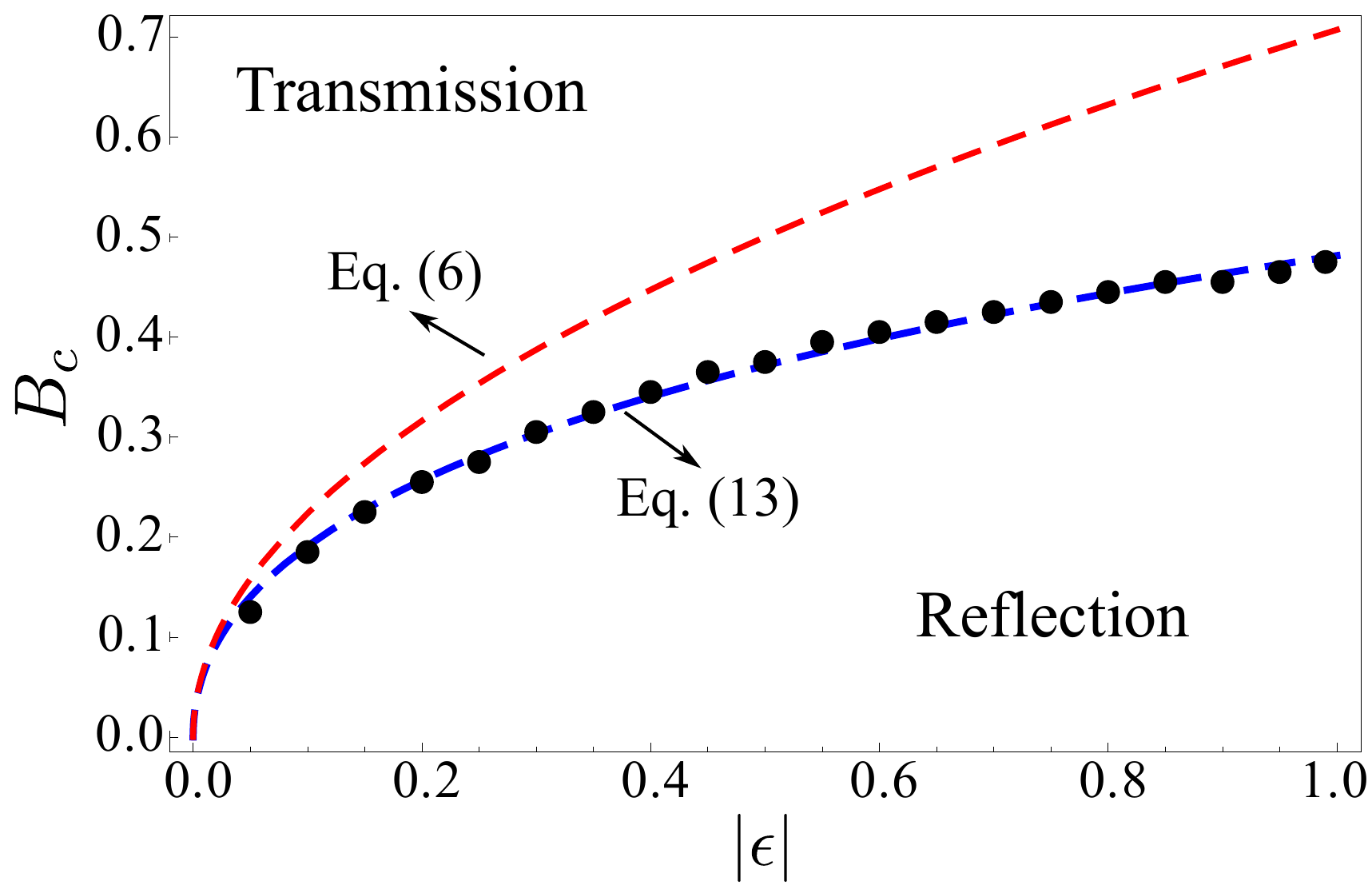}
\caption{Critical speed of the DDS ($B_c$) as a function of the strength of a negative
impurity, $\epsilon<0$. The dots denote the critical speed calculated numerically from Eq.~\eqref{eq-DNLS}, the dashed red line corresponds to the Hamiltonian estimation, Eq.~(6), and the 
dashed blue line is the fit given by Eq.~(13) with $\alpha=1.2$.}
\label{fig3}
\end{figure}
\section{Adiabatic approach}
To understand the DDS dynamics in the presence of impurities, we perform a continuous approach of the discrete system. For that, we introduce a phase shift of the
complex amplitude
$u_n(z)\rightarrow u(x_n,z)e^{2iz}$ in Eq.~\eqref{eq-DNLS}. And considering the discrete derivative as $(u_{n+1}+u_{n-1}-2u_{n})\propto
\left.\left(\partial^2 u/\partial x^2\right)\right|_{x=x_n}$, we obtain
\begin{equation}
\label{eq-CNLS}
-i\frac{\partial u}{\partial z} = \frac{\partial^2 u}{\partial x^2}
+V(x)u - |u|^2u\,,
\end{equation}
where a local impurity at the origin has been introduced as a delta-like potential,
$V(x) \equiv \epsilon\delta(x)$. To attain an analytical description of the evolution of the dark soliton (DS), we adopt the adiabatic perturbation theory
developed in Ref.~\cite{bec3} for dark solitons interacting with impurities in Bose-Einstein condensates. Hereby, we consider the complex
amplitude $u(x,z) = v(x,z)u_b(x)e^{-iz}$
in Eq.~\eqref{eq-CNLS}, where $u_b(x)=1+\frac{\epsilon}{2}e^{-2|x|}$
represents the wavefunction associated with the background field in the
presence of a delta impurity. Thus, the DS profile $v(x,z)$ satisfies the perturbed defocusing nonlinear Schr\"odinger equation
[see Ref.~\cite{bec3} and references therein]
\begin{equation}
\label{eq-SolProf}
i\frac{\partial v}{\partial z} +\frac{\partial^2v}{\partial x^2}-\left(|v|^2-1\right)v = P(v)\,,
\end{equation}
where $P(v) = \epsilon e^{-2|x|}[(1-|v|^2v)v - \text{sgn}(x)\partial_xv]$. It is known that in the unperturbed
limit ($P(v)=0$), i.e. the defocusing NLS equation, the DS solution
is given by $v(x,z) = \cos\phi\tanh\xi+i\sin\phi$, where $\xi = \cos\phi\left[x-(\sin\phi)z\right]$~\cite{ruso}. Then, following the
adiabatic treatment \cite{bec3}, after the introduction of the impurity, the parameters become slowly
varying functions of the propagation coordinate $z$ but the functional form remains unchanged.
Thus, the soliton phase is promoted to $\phi\rightarrow\phi(z)$ and the
soliton coordinate becomes $\xi\rightarrow \cos\phi(z)\left[x-x_0(z)\right]$.
Here, the variable $x_0(z)$ accounts for the core of the dark soliton. As a
result, the dynamics of the DS in the presence of an impurity is determined by
the Newtonian equation $d^2 x_0/d z^2 = -dU/dx_0$, with
$U(x_0) = -\epsilon/4\cosh^2(x_0)$. This force equation is
associated with the Hamiltonian
\begin{equation}
H =\frac{\dot{x}_0^2}{2}-\frac{\epsilon}{4\cosh^2(x_0)}\,.
\label{hamilton}
\end{equation}
In this case, $H_c=-\epsilon/4$ determines the transition for a DS to be reflected (when $H<H_c$) or
transmitted (for $H>H_c$) by the impurity. This defines a critical initial
speed given by
\begin{equation}
\label{eq-Bc}
B_c\approx \dot{x}_c = \pm\frac{\sqrt{-\epsilon}}{2}\,,
\end{equation}
which corresponds to the minimum speed that a dark soliton coming from infinity must have in order to be transmitted by the impurity.
Notice that here infinity is understood as an initial position at which the effect of the impurity over the DS can be neglected.
According to Eq.~\eqref{eq-Bc}, the transition arises only when $\epsilon<0$ in order to have real values of velocity ($\dot{x}_c\in \mathds{R}$), which
agrees with our previous numerical observations. Figure \ref{fig3} shows the comparison between the critical velocity for the DDS ($B_c$) and the Hamiltonian estimation given above.
We highlight that the critical speed obtained by the Hamiltonian description represents only a scaling approach. This is because continuous and discrete models are not exactly equivalent.
\begin{figure}[t]
\centering
\includegraphics[width=8.5cm]{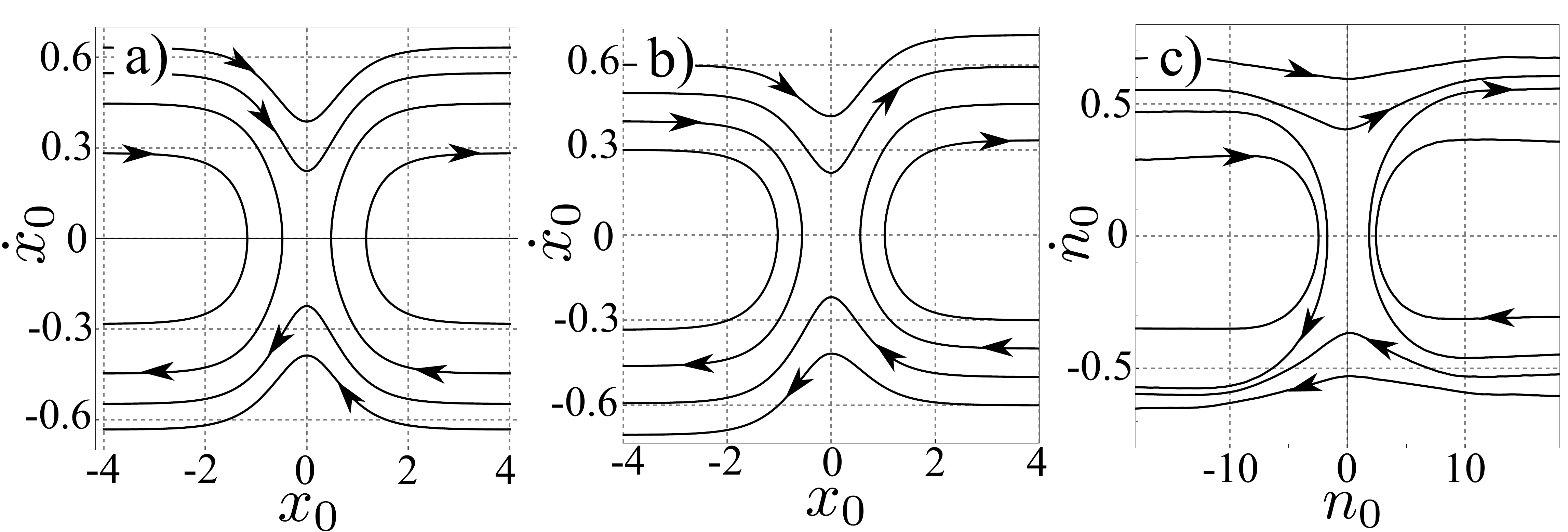}
\caption{Phase space for a system with one impurity. a) Elastic
desciption, b) inelastic description and c) numerical integration
of Eq.~\eqref{eq-DNLS}.}
\label{fig-PhaseSpace}
\end{figure}
\section{Inelastic description}
As expected, the Hamiltonian analysis [cf.~Eq.~\eqref{hamilton}]
leads to an elastic description of the scattering process. However, the
exchange of energy involves mainly three different type of waves: the
DDS, which can change its width and speed after each scattering. The
localized impurity mode, whose characteristic size depends on
$\epsilon$. And finally, radiation modes that spread over the background. Such interchange between different modes produces inelastic dynamics of scattering.
In order to confirm this, we construct the numerical phase space by extracting the speed and position of the center of the DDS from the simulation.
Since we have considered a local impurity, the speed change occurs in a
region close to the impurity. That is, the DDS travels with a constant speed until the interaction with the impurity takes place.
Consecutively, the speed of the DDS is increased and then, when the DDS is far enough from the impurity, the new speed remains constant
[cf.~Fig.~\ref{fig-diffdyn}]. Figure ~\ref{fig-PhaseSpace} shows the phase spaces for the Hamiltonian description, the numerical simulation, as well as, the inelastic model
that we will develop below. Numerical phase space in Fig. \ref{fig-PhaseSpace}c has been made by interpolating the DDS using a piecewise Hermite interpolation method. Then, we estimate the speed by tracking the minimum of the interpolated
function as a function of $z$.

In order to deliver a qualitative and quantitative description for the inelastic behavior, we introduce a phenomenological correction to the Newtonian equation
obtained after the adiabatic treatment. Based on our numerical observations discussed above: (i) the interaction is local and (ii) the speed of the DDS is not preserved.
The position of the soliton core can  thus be described by
\begin{equation}
\ddot{x}_0 = -\frac{dU(x_0)}{dx_0}(1+\bar{\alpha}(x_0)\dot{x}_0)\,,
\label{eq-Nonelastic}
\end{equation}
where $\bar{\alpha}(x_0)\equiv\alpha\cdot\text{sgn}(x_0)$. The constant parameter $\alpha$ accounts for the injection of energy due to the interaction with the impurity.
Note that the function $\text{sgn}(x_0)$ has been introduced to preserve the invariance of the system under spatial inversions ($x\to-x$) [cf.~Eq.~\eqref{eq-CNLS}].
Hereby, we are able now to determine the increment in the speed of a the DDS after being reflected by the impurity. We then perform a perturbative analysis
for the final speed of the DDS  at the left of the impurity approaching it with positive speed $\dot{x}_{f,0}$. It is important to mention that the final speed is evaluated at the $z$ where the DDS returns to its initial position, such that $x_i=x_f$.
Accordingly, we introduce the expansion $\dot{x}_f = \dot{x}_{f,0} + \alpha \dot{x}_{f,1}+
\alpha^2 \dot{x}_{f,2}+\cdots$ in Eq.~\eqref{eq-Nonelastic}, obtaining
\begin{eqnarray}
\ddot{x}_{f,0} &=& -\left.\frac{dU}{dx}\right|_{x=x_f}\,,\\
\ddot{x}_{f,n} &=&\left.\frac{dU}{dx}\right|_{x=x_f}\dot{x}_{f,n-1}=\frac{(-1)^{n}}{(n+1)!}\frac{d \dot{x}_{f,0}^{n+1}}{dz}\,.
\label{eq-X_Fn}
\end{eqnarray}
The former equation is the force equation obtained for the unperturbed (Hamiltonian) description, Eq.~\eqref{hamilton}.
Thus, we set $\dot{x}_{f,0}=-\dot{x}_i=-\sqrt{2}\sqrt{H-U(x_i)}$, where $\dot{x}_i$ is the initial speed and $H$ is the unperturbed initial energy.
Equation (9) leads to an asymptotic expansion of $\dot{x}_f$ in the form
\begin{equation}
\dot{x}_f = -x_i -\frac{\dot{x}_i^2}{2}\alpha - \cdots =
-\dot{x}_i\sum_{n=0}^{\infty}\frac{\dot{x}_i^n\alpha^n}{(n+1)!}\,.
\label{eq-VfDisc}
\end{equation}
This series can we written in  closed form as
\begin{equation}
\dot{x}_f=\frac{(1-e^{\alpha \dot{x}_i})}{\alpha}\equiv F(\dot{x}_i)\,.
\label{eq-VfCont}
\end{equation}
It is worth noting that the final speed reached by the DDS after being
reflected by the impurity, only depends on the speed that the DDS had before the interaction.
For a DDS approaching  the impurity from the right (with an initial speed $-\dot{x}_i$) we obtain a final
speed as in Eq.~\eqref{eq-VfDisc} but positive. In a compact form,
it reads $\dot{x}_f=(e^{\alpha \dot{x}_i}-1)/\alpha$.
Thus, a DDS colliding with the impurity from the left or the right will experience
the same increase in  speed as function of its initial speed. This is consistent with numerical simulations on the discrete system
[cf. Fig.~\ref{fig-PhaseSpace}].

\begin{figure}[t]
\centering
\includegraphics[width=8cm]{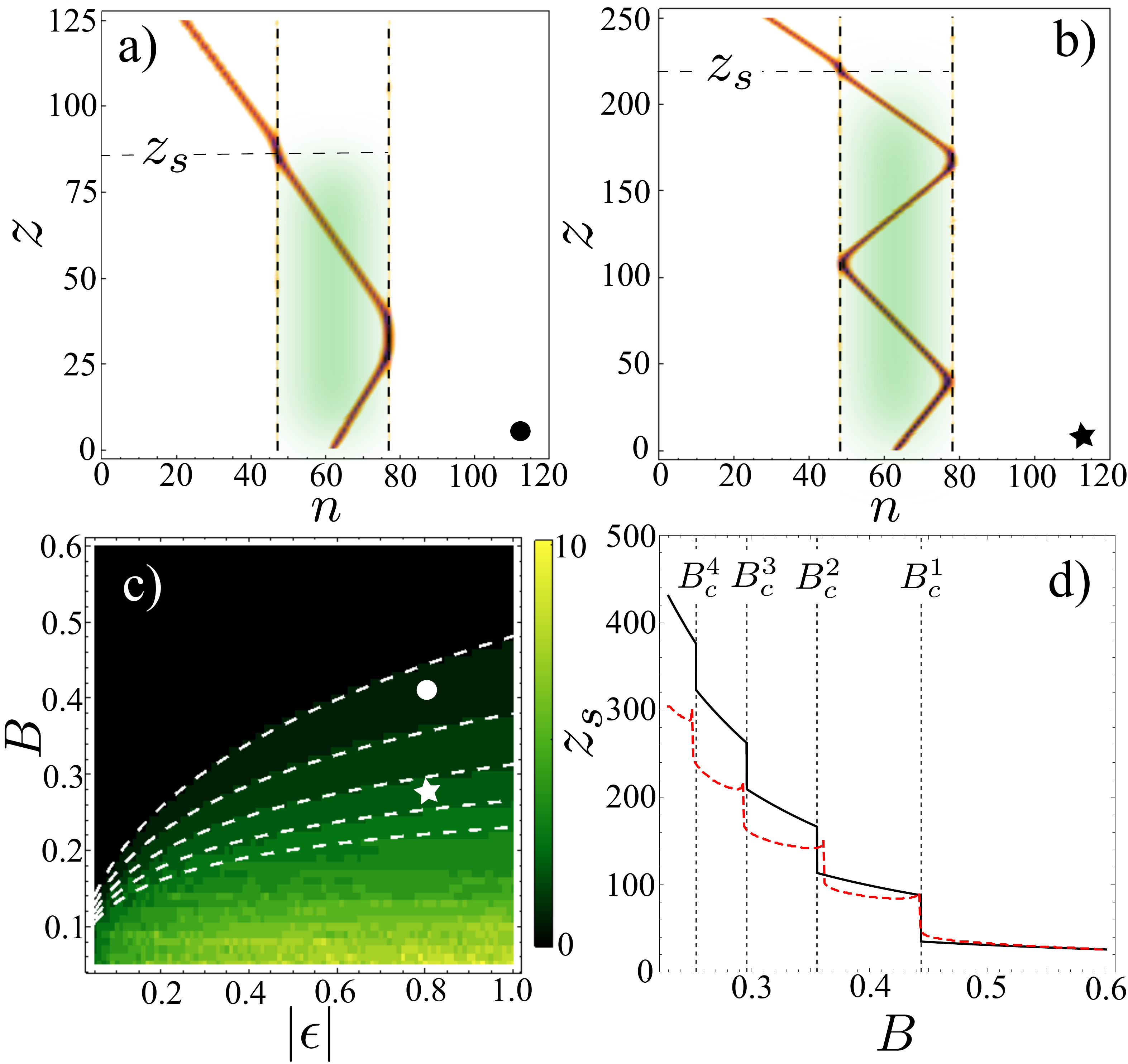}
\caption{a) and b) show the propagation of a DDS along the storage region (hatched in green) bounded by two
impurities (vertical dashed lines) for $B=0.4$ and $B=0.27$, respectively. The horizontal dashed lines mark the storage distance $z_s$.
c) Shows the number of reflections until the DDS is transmitted out of the storage region. White
dashed lines were calculated using Eqs. (12) and (13) for $\alpha = 1.2$. The
circle and the star mark the parameters used in a) and b), respectively. d)
Storage distance as function of $B$. Red dashed and black lines are associated with the numerical simulation and Eq. (14), respectively. In
a), b) and d) $\epsilon=-0.8$, and the impurities are separated by $30$ guides.}
\label{fig-TwoImp}
\end{figure}

\section{Storage mechanism}

The capability of negative impurities to reflect DDSs allows us to design a simple, but effective, mechanism of spatial storage. This consists of a system of two
impurities ($M=2$) distant enough such that the interaction of the DDS with each impurity occurs independently. Therefore, a DDS initially positioned
between impurities will remain in the storage region as long as its speed is less than the critical speed at which it is transmitted out
[see Fig.~\ref{fig-TwoImp}a,b]. Within the storage region, the DDS is consecutively reflected by the impurities delimiting the storage region. After each
collision the speed of the DDS increases according to Eq.~\eqref{eq-VfCont}. Note that the condition for the above to be valid is that the impurities are
sufficiently separated. Once the DDS reaches the critical speed to be transmitted by one of the the impurity, it will leave the storage region.
From Eq.~\eqref{eq-VfCont} we can build a recursive procedure to determine the number of reflections that a DDS will experience before reaching a
speed greater (or at least equal to) than the critical speed. Thus, we define the $n$-th critical speed
\begin{equation}
\label{eq-BcItera}
B_c^n \equiv \frac{\ln\left(1+\alpha B_c^{n-1}\right)}{\alpha}\,,
\end{equation}
as the maximum initial speed that a DDS can have such that
there will be $n$ reflections before escaping of the storage region.
That is, if the initial speed of the DDS is in the interval $(B_c^{n+1},B_c^n)$,
then it will be reflected $n$ times before being transmitted.
Note that Eq.~\eqref{eq-BcItera} requires one to know the minimal initial speed at which the DDS is transmitted. A
good estimation of this quantity is given by
\begin{equation}
B_c^1 =
\frac{\ln\left(1+\sqrt{-\epsilon}\right)}{\alpha^2}\,.
\end{equation}
Figure~\ref{fig-TwoImp}c show  good agreement between the analytical $n$-th critical velocity, Eq.~\eqref{eq-BcItera}, and the numerical critical
speed of the  DDS ($B_c$).

Finally, an upper value for the storage distance $z_s$ is estimated by considering that the speed reached by the DDS after scattering with one impurity,
remains constant until the next reflection
\begin{equation}
z_s =
d\left(-\frac{1}{2|\dot{x}_i|}+\sum_{m=1}^{N}\frac{1}
{\left|F^{(m)}(\dot{x}_i)\right|}\right), \;\text{if
}B_c^N<\dot{x}_i<B_c^{N-1}\,,
\label{eq-zs}
\end{equation}
where $d$ is the space between impurities. Fig.~\ref{fig-TwoImp}d compares the numerical and analytical results for a certain
storage distance. We can see that Eq.~\eqref{eq-zs} provides a good estimation for few reflections, however the error increases according to the number of reflections.

\subsection{Storage of staggered discrete dark solitons}
Throughout this article, all of our numerical simulations have involved the integration of Eq.~\eqref{eq-DNLS}, which intrinsically describes a discrete system.
Here, effects of discreteness in the propagation of the DDS gains importance mainly at very low speed, when the
DDS solutions are highly localized~\cite{dds}. Faster DDSs are well described by a continuous description.
Thus, we expect that storage behavior arises in continuous systems as well, described by the NLS equation in the presence of narrow impurities.
However, in this case dark solitons exist only in the self-defocusing regime, thus storage can be possible only in this nonlinear regime.

\begin{figure}[t]
\centering
\includegraphics[width=8cm]{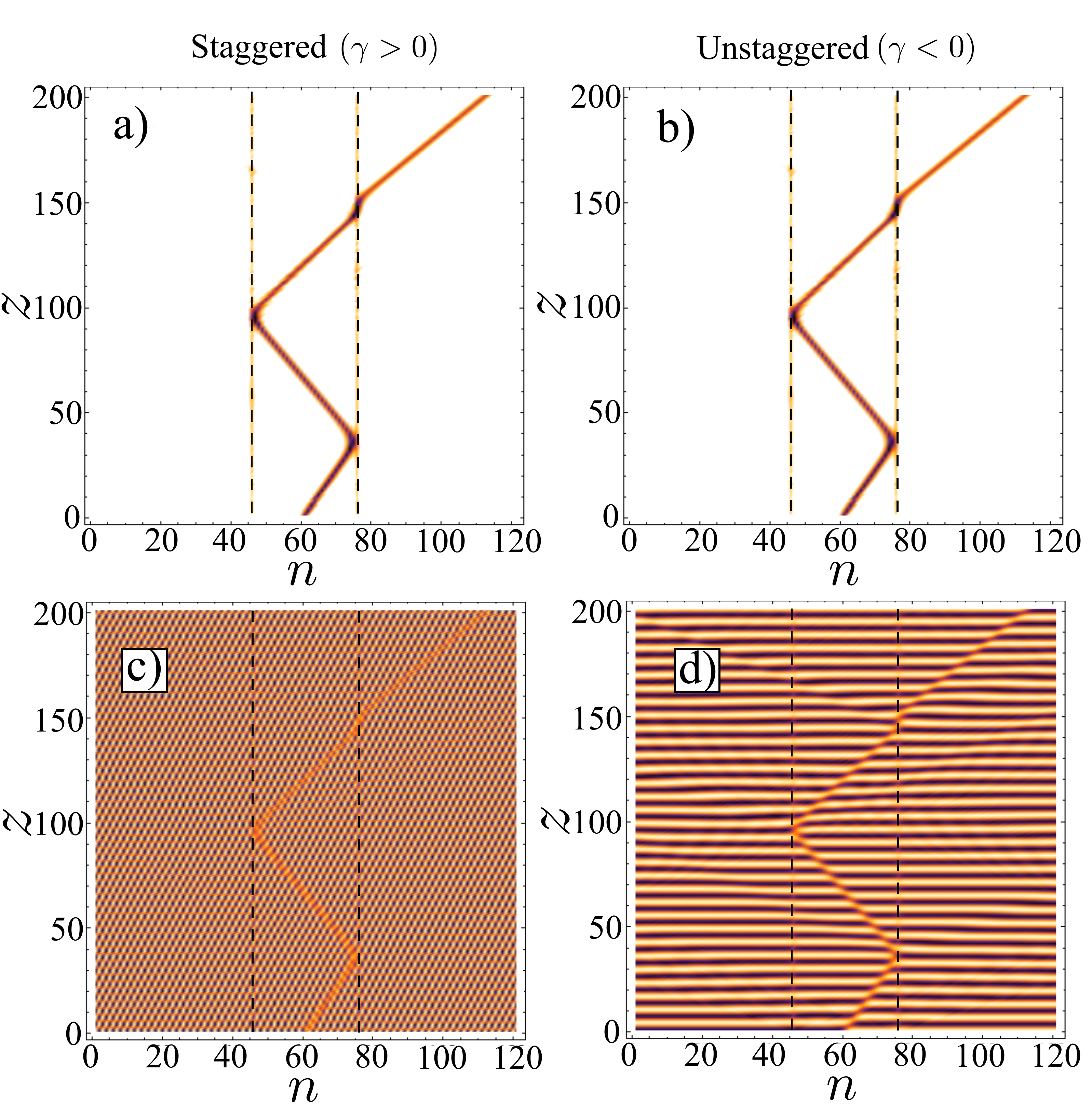}
\caption{Example of propagation of staggered (left panels, $\gamma=1$) and unstaggered (right panels, $\gamma=-1$) discrete dark solitons in a storage regime.
a) and b) represent the evolution of the intensity, while c) and d) the real part
of the optical field: $\text{Re}\{u_n\}$. Initial speed is given by $B=0.3$. The impurities are separated by $30$ guides, $|\epsilon|= 0.8$ and their positions are shown as vertical dashed lines.}
\label{fig-stg}
\end{figure}

Otherwise, in a waveguide configuration where DNLS applies, storage of DDS can be possible in both, self-defocusing and self-focusing nonlinear regimes, with appropriate phase engineering. In order to emphasize this, we compute numerical simulations in both cases and we did not find any difference
between either regime.
Of course, numerically this was expected since both regimes are mathematically equivalent because {\it staggered-unstaggered} transformation applies (see Sec. II).
Figure~\ref{fig-stg} shows
the intensity $|u_n|^2$ and the real part of the optical field $\text{Re}\{u_n\}$ associated with the dynamics of a DDS between two impurities. Here, we
can see that the only one difference between both fields
comes from the relative phase between neighboring guides, while the intensity distribution is the same.

\section{Storage of multiple DDS}

Now, an easy way to generate multiple discrete dark solitons is using as initial condition:
\begin{equation}
u_n(0) = 1 - C\sum_{k=1}^{K}\sech\left[D(n-n_k)\right]\,,
\end{equation}
which leads to the formation of $K$ pairs of DDSs that propagate with equal speed, but in opposite directions. Each pair is centered initially around $n_k$.
The speed can be controlled
by manipulating either $C\in[0,2]$ or $D>0$. One interesting fact about this initial condition, in contrast to the one used before, is that it requires a
homogeneous phase
distribution and only a spatial region (a few guides) where the intensity of the optical field decrease in comparison with its background value.

\begin{figure}[t]
\centering
\includegraphics[width=8cm]{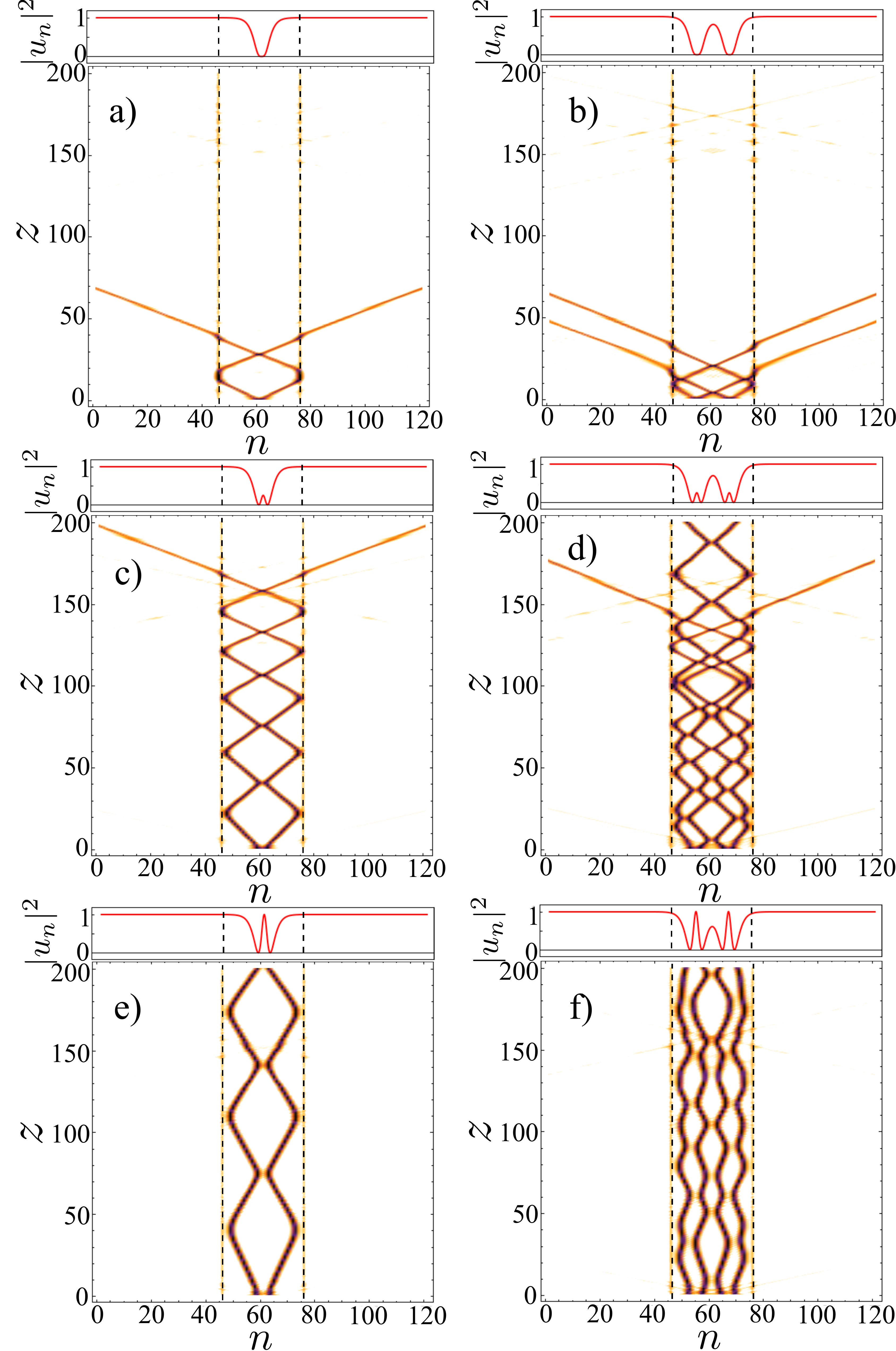}
\caption{Examples of propagation of multiple DDSs between two negative impurities. Left panels are associated with one pair of DDSs ($K=1$) centered at the middle of the impurities, while right panels for two
pairs of DDSs ($K=2$) separated by $n_2-n_1=12$ guides. The impurities are separated by $30$ guides and their positions are shown as vertical dashed lines. The parameters are: a)-b) $C=1$, c)-d) $C=1.5$, and e)-f) $C=2$. In each case $\epsilon=-0.8$ and $B=0.6$.
The initial intensity distribution $|u_n|^2$ is shown in the upper panels.}
\label{fig-MultiDDS}
\end{figure}

Numerically, we study the case with one ($K=1$) and two ($K=2$) pairs of DDSs initially placed between two negative impurities.
Some examples are shown in figure~\ref{fig-MultiDDS}, where
long-distance storage is observed ($z>400$). When $K=1$ and the initial configuration is symmetric with respect to the
central point between the two impurities,  the optical field will propagate symmetrically as well and finally the pair of DDSs escape simultaneously after
reaching the critical speed.
As in the case of one DDS, the mechanism is
the same and the DDS increases its speed after each inelastic interaction with the impurities. Furthermore, the storage distance depends only on the initial
speed and the strength of the impurities.
Nevertheless, the entire dynamic along $z$ is enriched by the elastic collisions undergo by the DDSs.

In the case of $K=2$, the internal dynamic is more complex and either periodic or aperiodic oscillations can be found, depending on the initial configuration.
Furthermore, if the initial condition
is symmetric, then the DDS will propagate symmetrically and DDSs will escape by pairs as long  as their speed increases enough. This is shown in
figure~\ref{fig-MultiDDS},
where we can see that faster DDSs escape, while slower DDSs are maintained within the storage region.

\section{Conclusions}

In conclusion, we have shown that the scattering of a DDS with a single
impurity is an inelastic process and the condition for
reflection or transmission  relies only on the
initial speed of the DDS and the strength of the impurity.
Furthermore, after the scattering, the speed of the DDS is increased.
A phenomenological description for the inelastic behavior of the DDS has been
proposed based on the well known adiabatic perturbation theory,
by including a non-Hamiltonian term, which accounts for the
acceleration of the DDS after each reflection with an impurity.

In addition, a method of spatial storage based on two negative
impurities is proposed. We show that a DDS can be trapped within a
storage region until it reaches the critical speed needed to be
transmitted. We showed that this method can be implemented successfully
for trapping multiple discrete dark solitons for a certain propagation distance.

\section*{Acknowledgements}

The authors are grateful to M. I. Molina and M. A. Garcia-\~Nustes
for useful discussions. A. J. M. and Y. Z. acknowledge the
support of CONICYT by BCH72130485/2013 and BCH721300436/2013,
respectively.

\bibliographystyle{unsrt}
\bibliography{References}

\end{document}